\documentclass[aps,pre,preprint,superscriptaddress,showpacs,showkeys]{revtex4-1}

\usepackage{amsmath, amsfonts, amsthm, amssymb, mathrsfs, enumerate, graphicx, cases, hyperref}
\usepackage[T1]{fontenc}

\begin{document}


\title{Mass-loss due to gravitational waves with $\Lambda>0$}


\author{Vee-Liem Saw}
\email[]{VeeLiem@maths.otago.ac.nz}
\affiliation{Department of Mathematics and Statistics, University of Otago, Dunedin 9016, New Zealand}


\date{\today}

\begin{abstract}
The theoretical basis for the energy carried away by gravitational waves that an isolated gravitating system emits was first formulated by Hermann Bondi during the 1960s. Recent findings from looking at distant supernovae revealed that the rate of expansion of our universe is accelerating, which may be well-explained by sticking in a positive cosmological constant into the Einstein field equations for general relativity. By solving the Newman-Penrose equations (which are equivalent to the Einstein field equations), we generalise this notion of Bondi mass-energy and thereby provide a firm theoretical description of how an isolated gravitating system loses energy as it radiates gravitational waves, in a universe that expands at an accelerated rate. This is in line with the observational front of LIGO's first announcement in February 2016 that gravitational waves from the merger of a binary black hole system have been detected.
\end{abstract}


\maketitle


The notion that gravitational waves carry energy away from an isolated system of masses was first established by Bondi and his coworkers, leading to the well-known mass-loss formula \cite{Bondi60,Bondi62}. This assumed that spacetime is asymptotically flat, i.e. the system of masses is confined within a bounded region and spacetime gets ever closer to being Minkowskian towards large distances away from the source. Bondi enunciated a metric ansatz describing an axially symmetric spacetime, and solved the vacuum Einstein equations (together with the Bianchi identities) in the region far away from the source. The mass-loss formula is then obtained from one of the ``supplementary conditions'' that arose from the Bianchi identities.

Around that same period in the 1960s, an equivalent formulation of general relativity was worked out by Newman and Penrose, making use of quantities called \emph{spin coefficients} (which are essentially the connection coefficients) \cite{newpen62}. This led to a collection of 38 (mostly linear) differential equations which are equivalent to the Einstein field equations and the Bianchi identities. By solving these equations at large distances for asymptotically flat spacetimes, Newman and Unti obtained the general asymptotic solutions \cite{newunti62}. One of the relationships from the Bianchi identities, integrated over a 2-sphere of constant $u$ at null infinity $\mathcal{I}$, is in fact the Bondi mass-loss formula
\begin{eqnarray}\label{Bondimasslossflat}
\frac{dM_B}{du}=-\frac{1}{A}\oint{|\dot{\sigma}^o|^2dS},
\end{eqnarray}
where $\displaystyle M_B=-\frac{1}{A}\oint{(\Psi^o_2+\sigma^o\dot{\bar{\sigma}}^o)}d^2S$ is the Bondi mass, and $A$ is the area of that 2-sphere of constant $u$ on $\mathcal{I}$. Dot is derivative with respect to $u$, which is a retarded null coordinate (and may be interpreted as ``time''). The term $\Psi^o_2$ is the leading order term of $\Psi_2$ when expanded over large distances from the source (where $\Psi_2$ is one of the dyad components of the Weyl spinor), and $\sigma^o$ is the leading order term of the complex spin coefficient $\sigma$ (under the large distance expansion). The presence of $\dot{\sigma}^o$ indicates gravitational waves being emitted by the system, so the mass of the system decreases due to energy carried away by those gravitational waves.

The purpose of this talk at the \emph{Conference on Cosmology, Gravitational Waves and Particles} is to present the key results of a generalisation to the above mass-loss formula for a universe with a positive cosmological constant $\Lambda>0$, as was recently reported in Ref. \cite{Vee2016}. This work is motivated by the fact that we have observed our universe to be expanding at an \emph{accelerated} rate \cite{cosmo1,cosmo2}, which spawned intense research over the past few years (see the introduction and list of references found in Ref. \cite{Vee2016}). A positive cosmological constant $\Lambda>0$ provides a simple explanation for the accelerated rate of expansion.

By solving the Newman-Penrose equations with a cosmological constant $\Lambda$ \`{a} la Newman-Unti \cite{newunti62} in Ref. \cite{Vee2016}, the mass-loss formula with $\Lambda>0$ (obtained from that same Bianchi identity as in the case for $\Lambda=0$) is:
\begin{eqnarray}
\frac{dM_{\Lambda}}{du}
&=&-\frac{1}{A}\oint{\left(|\dot{\sigma}^o|^2+\frac{\Lambda}{3}|\eth'\sigma^o|^2+\frac{2\Lambda^2}{9}|\sigma^o|^4+\frac{\Lambda^2}{18}\textrm{Re}(\bar{\sigma}^o\Psi^o_0)\right)d^2S}.\label{0BondimasslossGraonly}
\end{eqnarray}
Here, the definition of the mass with $\Lambda>0$ is proposed to be
\begin{eqnarray}\label{topsph}
M_{\Lambda}:=M_B+\frac{1}{A}\int{\left(\oint{\left(\Psi^o_2+\sigma^o\dot{\bar{\sigma}}^o\right)\frac{\partial}{\partial u}(d^2S)}\right)du}+\frac{\Lambda}{3A}\int{\left(\oint{K|\sigma^o|^2d^2S}\right)du},
\end{eqnarray}
which ensures that this mass $M_\Lambda$ strictly decreases whenever $\sigma^o$ is non-zero (in the absence of incoming gravitational radiation from elsewhere, or $\Psi^o_0=0$), i.e. the mass of an isolated gravitating system \emph{strictly decreases due to energy carried away by gravitational waves} \cite{Vee2016}. (The term with $K$ in Eq. (\ref{topsph}) will be explained below, shortly.)

As opposed to the case where $\Lambda=0$, here $\mathcal{I}$ is \emph{non-conformally flat} for $\Lambda>0$ --- its Cotton-York tensor is non-zero, when those outgoing gravitational waves carry energy away from the source \cite{ash1,Vee2016}. As a result, the integration is not carried out over a round 2-sphere --- but instead over a \emph{topological 2-sphere} of constant $u$ on $\mathcal{I}$ \cite{Szabados,Vee2016} \footnote{See Ref. \cite{Szabados} which describes the foliation of the conformally rescaled $\mathcal{I}$ by topological 2-spheres.}. These compact 2-surfaces of constant $u$ on $\mathcal{I}$ being topological instead of round 2-spheres give rise to the second term on the right-hand side of Eq. (\ref{topsph}), because the surface element $d^2S$ in general has a $u$-dependence \footnote{This crops up due to a process of interchanging the order of taking a $u$-derivative and integrating over the topological 2-sphere.}. The Gauss curvature for these topological 2-spheres on $\mathcal{I}$ is $K$, which in general depends on $\sigma^o$ when $\Lambda\neq0$. If $\sigma^o=0$, then $K=1$ --- indicating a round 2-sphere of constant curvature.

We see that with the presence of $\Lambda$, then $\sigma^o$ itself would contribute to the mass-loss formula in Eq. (\ref{0BondimasslossGraonly}). (Recall that for asymptotically flat spacetimes in Eq. (\ref{Bondimasslossflat}), one needs a variation of $\sigma^o$ with $u$ to give rise to a decrease in mass.) Apart from that, $\Psi^o_0$ (which is the leading order term of the dyad component of the Weyl spinor $\Psi_0$) represents incoming radiation and shows up in the mass-loss formula --- due to $\mathcal{I}$ being a spacelike hypersurface. In a universe expanding at an accelerated rate, the isolated system has a cosmological horizon. Such incoming radiation from elsewhere that lies beyond the cosmological horizon would not affect the system, but \emph{will arrive at $\mathcal{I}$ and get picked up by the mass-loss formula}.

The asymptotic solutions with a cosmological constant have been extended to include Maxwell fields \cite{Vee2017}. The full mass-loss formula is
\begin{eqnarray}
\frac{dM_{\Lambda}}{du}
&=&-\frac{1}{A}\oint{\left(|\dot{\sigma}^o|^2+k|\phi^o_2|^2+\frac{\Lambda}{3}|\eth'\sigma^o|^2+\frac{2\Lambda^2}{9}|\sigma^o|^4+\frac{\Lambda^2}{18}\textrm{Re}(\bar{\sigma}^o\Psi^o_0)-\frac{k\Lambda^2}{36}|\phi^o_0|^2\right)d^2S},\ \ \label{0Bondimasslosspsi0ST}
\end{eqnarray}
and if there are only Maxwell fields but no gravitational radiation, i.e. $\sigma^o=0$, then
\begin{eqnarray}
\frac{dM_{\Lambda}}{du}
&=&-\frac{k}{A}\oint{\left(|\phi^o_2|^2-\frac{\Lambda^2}{36}|\phi^o_0|^2\right)d^2S},\label{0BondimasslossMaxonly}
\end{eqnarray}
where $M_\Lambda=M_B$ (since the compact 2-surface of constant $u$ on $\mathcal{I}$ is a round sphere when $\sigma^o=0$, so the surface element $d^2S$ does not have a $u$-dependence). The terms $\phi^o_2$ and $\phi^o_0$ (which are leading order terms of the dyad components of the Maxwell spinor $\phi_2$ and $\phi_0$, respectively) represent outgoing and incoming electromagnetic radiations, respectively. The Maxwell fields do not affect the structure of $\mathcal{I}$, unlike the gravitational counterpart which would lead to its non-conformal flatness when outgoing gravitational waves carry energy away from the isolated system. From Eq. (\ref{0BondimasslossMaxonly}), it is clear that outgoing electromagnetic radiation that the source emits carries energy away from it, whilst incoming electromagnetic radiation from elsewhere would increase the isolated system's total mass-energy.

Full details and extensive elaborations are given in Ref. \cite{Vee2016} for the gravitational case, with the extension to include Maxwell fields found in Ref. \cite{Vee2017}.

\begin{acknowledgments}
V.-L. Saw is working on this research project under the University of Otago Doctoral Scholarship. Travel support to Nanyang Technological University (NTU), Singapore for this Conference was provided by the Division of Science, University of Otago. Furthermore, I am very grateful for the hospitality shown by the Institute of Advanced Studies, NTU in hosting my stay throughout the Conference.
\end{acknowledgments}

\bibliographystyle{spphys}       
\bibliography{Citation}

\end{document}